\documentclass[useAMS,usenatbib]{mn2e}
\usepackage{graphicx}

\title[Rapid Binary Stellar Population Synthesis]{An Isochrone Database and a Rapid Model for Stellar Population Synthesis
\thanks{All the data are available at the CDS or on request to the authors.}}
\author[Zhongmu Li and Zhanwen Han]{Zhongmu Li$^{1,2}$ \thanks{E-mail:
zhongmu.li@gmail.com} and Zhanwen Han$^{1}$\\
$^{1}$National Astronomical Observatories/Yunnan Observatory, the
Chinese Academy of Sciences, Kunming, 650011,
   China\\
$^{2}$Graduate School of the Chinese Academy of Sciences}
\begin{document}

\date{Accepted 1988 December 15. Received 1988 December 14; in original form 1988 October 11}

\pagerange{\pageref{firstpage}--\pageref{lastpage}} \pubyear{2002}

\maketitle

\label{firstpage}

\begin{abstract}
We first presented an isochrone database that can be widely used for
stellar population synthesis studies and colour-magnitude diagram
(CMD) fitting. The database consists of the isochrones of both
single star and binary star simple stellar populations (ss-SSPs and
bs-SSPs). The ranges for the age and metallicity of populations are
0--15 Gyr and 0.0001--0.03, respectively. All data are available for
populations with two widely used initial mass functions (IMFs),
i.e., Salpeter IMF and Chabrier IMF. The uncertainty caused by the
database (about 0.81\%) is designed to be smaller than those caused
by the Hurley code and widely used stellar spectra libraries (e.g.,
BaSeL 3.1) when the database is used for stellar population
synthesis.

Then based on the isochrone database, we built a rapid stellar
population synthesis ($RPS$) model and calculated the
high-resolution (0.3 $\rm {\AA}$) integrated spectral energy
distributions (SEDs), Lick indices and colour indices for bs-SSPs
and ss-SSPs. In particular, we calculated the $UBVRIJHKLM$ colours,
$ugriz$ colours and some composite colours that consist of
magnitudes on different systems. These colours are useful for
disentangling the well-known stellar age--metallicity degeneracy
according to our previous work.

As an example for applying the isochrone database for CMD fitting,
we fitted the CMDs of two star clusters (M67 and NGC1868) and
obtained their distance moduli, colour excesses, stellar
metallicities and ages. The results showed that the isochrones of
bs-SSPs are closer to those of real star clusters. It suggests that
the effects of binary interactions should be taken into account in
stellar population synthesis. We also discussed on the limitations
of the application of the isochrone database and the results of the
$RPS$ model.
\end{abstract}


\section{Introduction}
Stellar population synthesis is a widely used technique to model the
spectra and photometry evolution of galaxies. It is also an
important method to estimate the stellar contents of galaxies from
spectra (e.g., \citealt{Trager:2000}) or photometry data (e.g.,
\citealt{Li:2007potential}; \citealt{Li:2007effects}). A lot of
models such as \cite{Bruzual:2003} (GALAXEV), \cite{Worthey:1994},
\cite{Vazdekis:1999} , \cite{Fioc:1997} (PEGASE), and
\cite{Zhang:2005} have been brought forward for studying stellar
populations, but most of them did not take binary interactions into
account. Similarly, the popular models for galaxy formation and
evolution (e.g., \citealt{Baugh:1998}, \citealt{Cole:2000},
\citealt{DeLucia:2006}) did not take binary interactions into
account, either. Therefore, most results of stellar population
studies are derived from single star stellar population (ss-SSP)
models, e.g., \citet{Terlevich:2002}, \citet{Gallazzi:2005},
\citet{Trager:2000}, while only a few results from binary star
stellar population (bs-SSP) model \citep{Li:2006bpsstudy}. A reason
for those works using ss-SSPs rather than bs-SSPs is that the
evolution of binaries is much more complicated than single stars.
The calculation of binary stellar population synthesis usually takes
much more time and disk space compared to single stellar population
synthesis. However, as pointed out by, e.g., \citet{Han:2001},
\citet{Zhang:2004}, more than 50\% stars are in binaries, and binary
interactions are important for stellar population synthesis. Many
observational phenomena such as the Far-UV excess of elliptical
galaxies \citep{Han:2007} and blue stragglers in star clusters
(\citealt{Tian:2006}; \citealt{Xin:2007}) can be explained more
naturally via bs-SSPs than via ss-SSPs. In fact, both blue
stragglers and Far-UV excess of elliptical galaxies can be produced
naturally by binary interactions without any special assumptions
(see \citealt{Han:2007} for more details). This suggests that it is
necessary to model stellar populations of galaxies via bs-SSPs. We
mainly intend to build a database for stellar population synthesis
studies and present a new model for quickly modeling bs-SSPs.

The structure of the paper is as follows. In Sect. 2, we briefly
introduce the evolution of stars. In Sect. 3, we present the
isochrone database for stellar population synthesis studies. In
Sect. 4, we present our new stellar population synthesis model. As
an application of the isochrone database in colour-magnitude diagram
(CMD) studies, we fit the CMDs of two star clusters in Sect. 5.
Finally, we give our discussions and conclusions in Sect. 6.

\section{Evolution of stars}

In this work, we use the rapid stellar evolution code of
\citet{Hurley:2002}, hereafter Hurley code, to evolve binaries and
single stars. This code enables modeling of even the most complex
binary systems. Binary interactions such as mass transfer, mass
accretion, common-envelope evolution, collisions, supernova kicks,
angular momentum loss mechanism, and tidal interactions are
considered (see \citealt{Hurley:2002} for details). Besides the code
can calculate the evolution of stars quickly, the average
uncertainty caused by the code is typically smaller than 5\%.

\section{The isochrone database}

\subsection{Input parameters}

To build a database for conveniently
and quickly modeling stellar populations, we take wide ranges for
input stellar-population parameters (metallicity, $Z$; age, $t$;
initial mass function, IMF). In detail, 0.0001--0.03 and
0--15 Gyr are taken for the ranges of $Z$ and $t$, respectively, and two widely used IMFs
are taken in the work. The two IMFs are presented by
\citet{Salpeter:1955} and \citet{Chabrier:2003} respectively and are listed
in Table 1.
\begin{table*}
 \caption{The IMFs adopted in this work.}
 \label{symbols}
 \begin{tabular}{lll}
  \hline
  \hline
   IMF name &IMF expression  &Example models \\
 \hline
    Salpeter &$\phi(m) \propto$
       $m^{-2.35}$ &
      {\rm \citet{Worthey:1994},  \citet{Zhang:2005}}\\
  \\
  \\
    Chabrier & $\phi({\rm log}~m) \propto  \left\{
     \begin{array}{ll}
       \exp[{{(\log m - \log m_{\rm c})^2}\over{-2\sigma^2}}],~~(m \leq 1M_\odot; ~m_{\rm c} = 0.08 M_\odot, \sigma = 0.69)\\
        \\
       m^{-1.3}~~(m >   1M_\odot)
     \end{array}
   \right.$
   &~~\citet{Bruzual:2003}\\
 \hline
\\
\multicolumn{2}{l}{{\it Note}: The mass of each star is given between 1 and 100 $M_\odot$ in this work, as \citet{Bruzual:2003}.}\\
 \end{tabular}
 \end{table*}

We use a Monte Carlo method to generate the binary sample of
bs-SSPs. For each binary, we generate the masses of two components
($M_{\rm 1}$ and $M_{\rm 2}$), the separation between the two
components ($a$), and the eccentricity of the binary ($e$). For
ss-SSPs, we just evolve stars independently. Therefore, the bs-SSPs
and ss-SSPs have the same star sample. We take the same distribution
as \citet{Mazeh:1992} and \citet{Goldberg:1994} for $q$, and a
distribution as \citet{Han:1995} for $a$. The detailed process for
generating the input parameters is as follows.

First, we generate the mass of the primary, $M_{\rm 1}$,
within the range of 0.1--100 $M_\odot$, according to an IMF.
Next we generate the ratio ($q$) of the mass of the secondary to
that of the primary randomly within 0--1, due to an a
uniform distribution
\begin{equation}
p(q) = 1,  ~~0 \leq q \leq 1,
\end{equation}
where $q = M_{\rm 2}/M_{\rm 1}$.
Then the mass of the secondary star is given by
$q$.$M_{\rm 1}$. One can also refer to \citet{Zhang:2004}.

Second, we generate the separation ($a$) of two components of a binary
following the assumption that the fraction of binary in an interval of log($a$) is constant
when $a$ is big (10$R_\odot$ $< a <$ 5.75 $\times$ 10$^{\rm 6}$$R_\odot$)
and it falls off smoothly when when $a$ is small ($\leq$ 10$R_\odot$).
The distribution of $a$ can be written as
\begin{equation}
  a~.p(a) = \left\{
            \begin{array}{ll}
            a_{\rm sep}(a/a_{\rm 0})^{\psi}, &~a \leq a_{\rm 0}\\
            a_{\rm sep}, &~a_{\rm 0} < a < a_{\rm 1}\\
     \end{array}
    \right.
\end{equation}
where $a_{\rm sep} \approx 0.070, a_{\rm 0} = 10R_{\odot}, a_{\rm 1}
= 5.75 \times 10^{\rm 6}R_\odot$ and $\psi \approx 1.2$. This
distribution implies that about 50\% (the typical value of the
Galaxy, see  \citealt{Han:1995}) of stars are in binaries with
orbital periods less than 100 yr.

Third, we generate the eccentricity ($e$) of each binary system
using a uniform distribution, in the range of 0--1, as
\citet{Zhang:2004}. Actually, the distribution of $e$ affects the
stellar population synthesis results slightly, according to
\citet{Zhang:2004}.

 \subsection{Building of the database}

To build an useful database for stellar population synthesis, the
isochrone database has to contain all the data needed in stellar
population synthesis. Three stellar-evolution parameters (effective
temperature, $T_{\rm eff}$, surface gravity, log($g$), and
luminosity, log($L/L_{\odot}$)) of stars in populations are
important for stellar population synthesis and we include them into
the database. In the work, each stellar population consists of
2\,000\,000 binaries or 4\,000\,000 single stars. In fact, the
sample contains two times of stars in the sample of
\citet{Zhang:2005} and some satisfying population synthesis results
can be obtained via such a sample (see \citealt{Zhang:2005} for more
details). To make the library can be used combining with most widely
used spectra libraries to calculate the integrated specialities such
as spectral energy distributions (SEDs) and magnitudes of stellar
populations, we take wide ranges for the stellar-evolution
parameters. The ranges are wider than those of most stellar spectra
libraries, e.g., BaSeL 3.1 \citep{Westera:2002}, STELIB
\citep{LeBorgne:2003}, and the library of \citet{Delgado:2005}. In
detail, our results are presented within the range of 2\,000 --
60\,000 K for $T_{\rm eff}$, and -1.5 -- 6 for log($g$).

To save disk space, the stellar-evolution parameters of stellar
populations are given by a statistical method. In other words, the
database supplies us with approximate distributions of stars in the
log($g$) versus $T_{\rm eff}$ grid (hereafter $gT$-grid), i.e.,
approximate isochrones, rather than the stellar-evolution parameters
of each star. The detailed procedure is as follows. First, we divide
$gT$-grid into 1\,089\,701 sub-grids, with intervals of 0.01 and 40
K for log($g$) and $T_{\rm eff}$, respectively. The two intervals
lead to an average uncertainty of 0.81\% in stellar population
synthesis results, which is smaller than those caused by the Hurley
code (5\%) and most stellar spectra libraries (e.g., 3--5\% for
BaSeL 3.1). Therefore, our results is accurate enough for most
stellar population synthesis studies. Second, for each stellar
population, we count the stars locate in each sub-grid, and save the
median log($g$), median $T_{\rm eff}$, average log($L/L_{\odot}$)
and percentage of stars in the sub-grid. The reason for saving the
median values rather than the average values of log($g$) and $T_{\rm
eff}$ is that the two kinds of values lead to almost the same
stellar population synthesis results but each pair of median
log($g$) and $T_{\rm eff}$ corresponds to a fixed sub-grid of the
$gT$-grid and they can be used more conveniently in stellar
population synthesis studies. The four parameters can describe the
isochrones and Hertzsprung--Russell diagrams (HRDs) of stellar
populations. They can also be used for calculating the average
surface area of stars in each sub-grid.

\subsection{The database}
As a whole, the isochrone database contains the distributions of
$T_{\rm eff}$, log($g$), and log($L/L_{\odot}$) of stars, i.e.,
isochrones of bs-SSPs and ss-SSPs. The age range for the database is
0--15 Gyr, with an interval of 0.1 Gyr, and the metallicity range is
0.0001--0.03 (0.0001, 0.0003, 0.001, 0.004, 0.01, 0.02 and 0.03). In
special, the data for stellar populations with Salpeter and Chabrier
IMFs are presented. The database can be used to model the CMDs of
star clusters and calculate the integrated specialities of stellar
populations. The database can be obtained by on request to the
authors or via the CDS in the future.

To understand the database more expediently, we show the isochrones
of metal-poor ($Z$=0.0001), solar-metallicity ($Z$=0.02) and
metal-rich ($Z$=0.03) stellar populations in Figs. 1, 2 and 3,
respectively. The populations here have Salpeter IMF. Note that we
take stellar populations with Salpeter IMF as our standard models
and we only show the results for standard models in the whole paper.
In the figures, both the isochrones of bs-SSPs and ss-SSPs are
shown, which can help us to understand the effects of binary
interactions on stellar population synthesis. We see clearly that
the isochrones of bs-SSPs are different significantly from those of
ss-SSPs. We will find that the isochrones of bs-SSPs are closer to
those of star clusters when we fit the CMDs of two star clusters
later.

\begin{figure}
  \includegraphics[angle=-90,width=88mm]{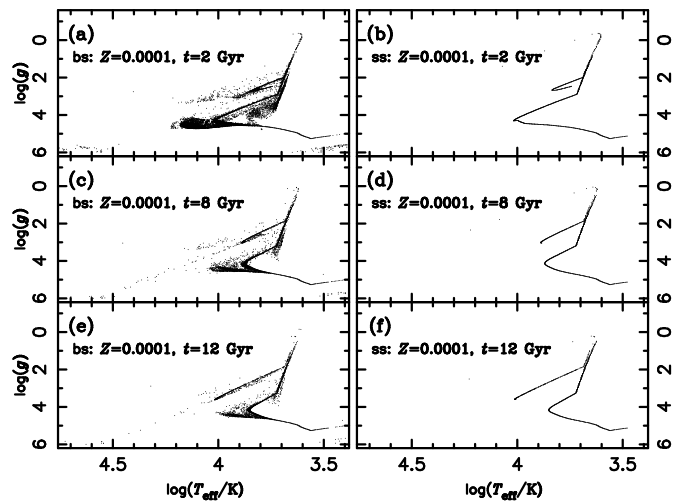}
  \caption{Isochrones for binary stellar populations (bs-SSPs) and single stellar populations (ss-SSPs)
  with metallicity of 0.0001 and Salpeter IMF. Symbols ``bs'' and
  ``ss'' denote bs-SSP and ss-SSP, respectively.
  The points show only the range of the distribution rather than the number of stars.}
\end{figure}

\begin{figure}
  \includegraphics[angle=-90,width=88mm]{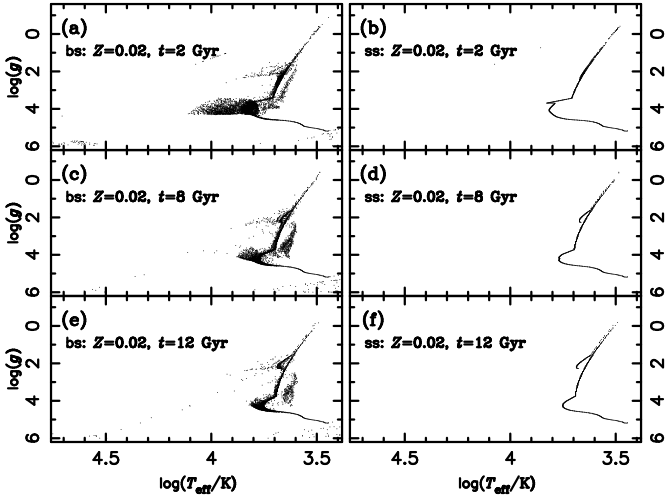}
  \caption{Similar to Fig.1, but for stellar populations with metallicity of 0.02.}
\end{figure}

\begin{figure}
  \includegraphics[angle=-90,width=88mm]{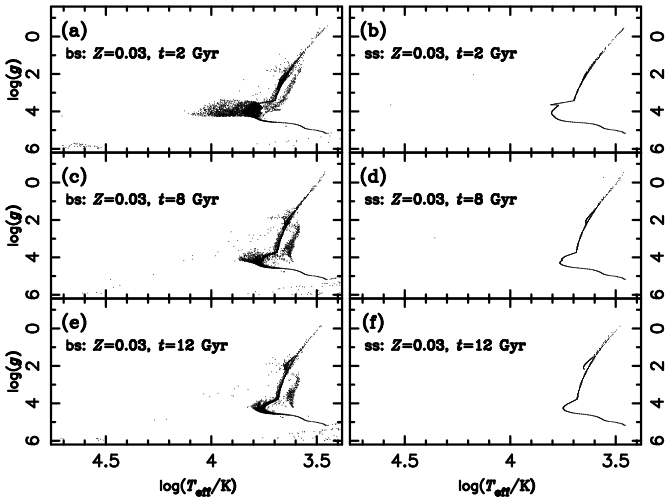}
  \caption{Similar to Fig.1, but for stellar populations with metallicity of 0.03.}
\end{figure}

\section{Rapid population synthesis model}

The isochrone database above enables quickly calculating the
spectral energy distributions (SEDs) of stellar populations, and
then the Lick Observatory Image Dissector Scanner indices (Lick
indices) and colour indices. Besides the method does not need to
evolve each star, it computes only one time for the spectra of stars
in each sub-grid of the $gT$-grid. Therefore, via such technique, we
can calculate the integrated peculiarities of stellar populations
very quickly. We call this stellar population synthesis technique
and model \emph{rapid stellar population synthesis ($RPS$)}. It
actually  takes much less (about one of 100\,000) time than the
method used by, e.g., \citet{Zhang:2005}. Therefore, the $RPS$
method makes it more convenient to take binary interactions into
account when modeling the formation and evolution of galaxies. In
the work, we calculated the SEDs, Lick indices and colour indices
for bs-SSPs and ss-SSPs. They can be conveniently used for future
studies.

\subsection{SEDs and Lick indices}
We calculated the SEDs and 25 widely used Lick indices for bs-SSPs
and ss-SSPs with two IMFs. The high-resolution stellar spectra
library of \citet{Martins:2005} (hereafter Martins library) was used
for the work. The library was computed with the latest results of
stellar atmospheres studies and covers a $T_{\rm eff}$ range from
3\,000 to 55\,000 K and a log($g$) range from -0.5 to 5.5. In
particular, the library enables modeling the spectra of stellar
populations at a 0.3 ${\rm \AA}$ resolution. Thus the library is
suitable for spectral studies of stellar populations of galaxies and
star clusters. However, the metallicity coverage of the library is
only from 0.002 to 0.04. In Figs. 4, 5 and 6, we show the SED
evolution of bs-SSPs with metallicities of 0.004, 0.02 and 0.03,
respectively. As we see, as that of ss-SSPs (see, e.g.,
\citealt{Bruzual:2003}), the continuum trends to be redder with
increasing age of bs-SSP and the metallicity of bs-SSP affects the
metal line (e.g., Fe lines with central wavelengths of 5270 and 5335
${\rm \AA}$) strengths obviously. We can also check the effects of
stellar age and metallicity via some Lick indices. In Figs. 7, 8, 9
and 10, we plotted the evolution of four widely used Lick indices.
The indices shown are computed from SEDs with a 0.3 ${\rm \AA}$
resolution, but we also calculated the indices on the Lick system.
Note the work calculates the Lick indices of populations from SEDs,
but Zhang et al.'s works calculate the indices on the Lick system
using some fitting functions. The figures (7, 8 and 9) show that the
stellar metallicity affects the age-sensitive line index H$\beta$
slightly, while it affects metallicity-sensitive indices (Mgb,
Fe5270, and Fe5335) strongly, and the stellar age changes the
H$\beta$ index similarly for populations with various metallicities.
Therefore, the line indices of bs-SSPs have similar age and
metallicity sensitivities as those of ss-SSPs. This suggests that
when we take bs-SSP models instead of ss-SSP models to study the
stellar populations of galaxies, we can use H$\beta$ together with
[MgFe] \citep{Thomas:2003} to give estimates for the age and
metallicity of populations. To check the reliability of our $RPS$
technique, we also plotted the results derived by \citet{Zhang:2005}
in Figs. 7, 8, 9 and 10. The two models took the same IMF, stellar
evolution code and spectra library, and both of them are bs-SSP
models, but their age ranges, star samples and computing method are
different. It is shown that the $RPS$ model gave line indices
similar to those of the previous work, but the indices obtained by
this work evolve more smoothly. The reason is that we take
2\,000\,000 binaries for a bs-SSP in the work but it is 1\,000\,000
in the previous work. Note that we obtained different metal indices
for bs-SSPs with metallicity of 0.01. The reason is as follows. The
Martins library supplies the same set of spectra for stars with
metallicities ([$Z$/H]) of -0.3 and -0.5. We took the set of spectra
as the spectra of stars with [$Z$/H] = -0.3, but the same set of
spectra were taken as the spectra of stars with [$Z$/H] = -0.5 in
the work of \citet{Zhang:2005}, when calculating the SEDs of
populations with metallicity ($Z$) of 0.01. The good consistency of
our results with those of the previous work suggests that our
isochrone database can be used reliably for stellar population
studies. Furthermore, we find that the abrupt change of indices
between 1 and 2 Gyr are possibly caused by the 'phase' transitions
of stars with initial mass between about 1.57 and 2 $M_{\rm \odot}$.
Such stars are in core helium burning stage and have high
luminosities between 1 and 2 Gyr. As a result, it leads to a
irregularity in the line indices and colours (see
\citealt{Bruzual:2003} for comparison). However, the values of
indices in the age range are possibly affected by the rough
calculation of the evolution of stars by the Hurley code.

\begin{figure}
  \includegraphics[angle=-90,width=88mm]{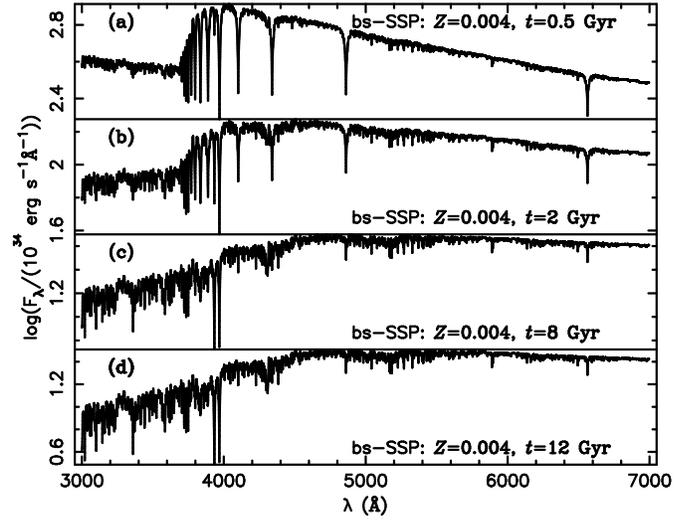}
  \caption{Evolution of spectral energy distributions (SEDs) of metal-poor ($Z$ = 0.004) bs-SSPs.
   Note that the y-axes of four panels are valued independently.}
\end{figure}

\begin{figure}
  \includegraphics[angle=-90,width=88mm]{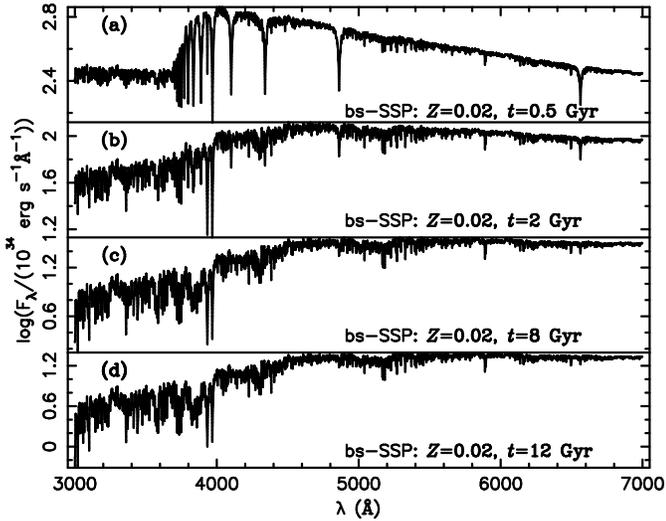}
  \caption{Similar to Fig.4, but for solar-metal bs-SSPs.}
\end{figure}

\begin{figure}
  \includegraphics[angle=-90,width=88mm]{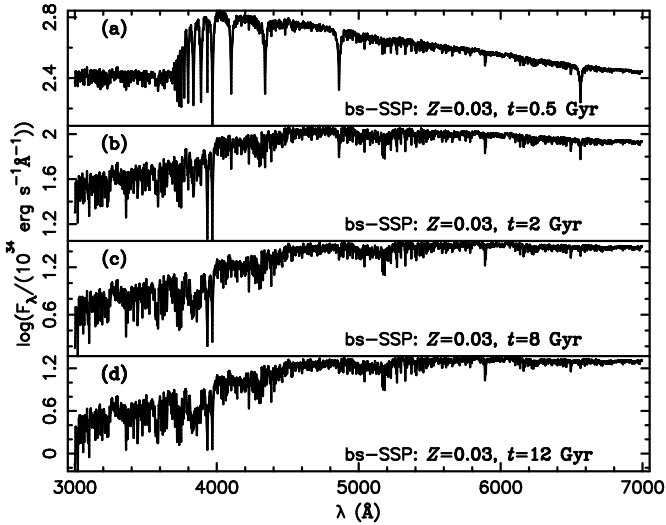}
  \caption{Similar to Fig.4, but for metal-rich ($Z$ = 0.03) bs-SSPs.}
\end{figure}

\begin{figure}
  \includegraphics[angle=-90,width=88mm]{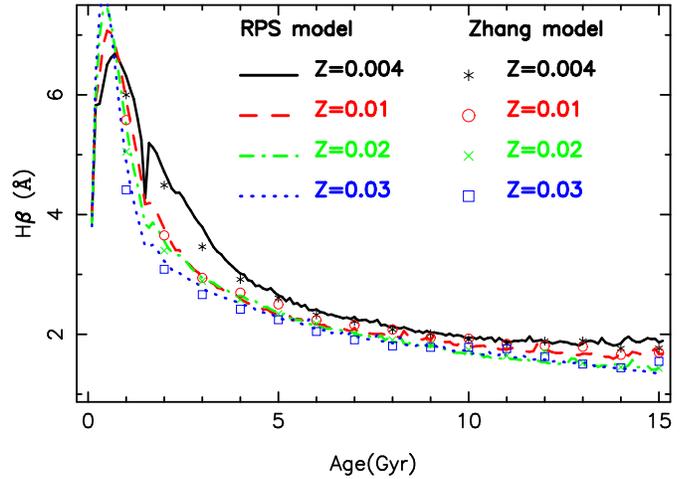}
  \caption{Comparison of the H$\beta$ index obtained by this work with
those by \citet{Zhang:2005}. The indices are computed from SEDs
directly. $RPS$ indicates our model. The two models are bs-SSP
model.}
\end{figure}

\begin{figure}
  \includegraphics[angle=-90,width=88mm]{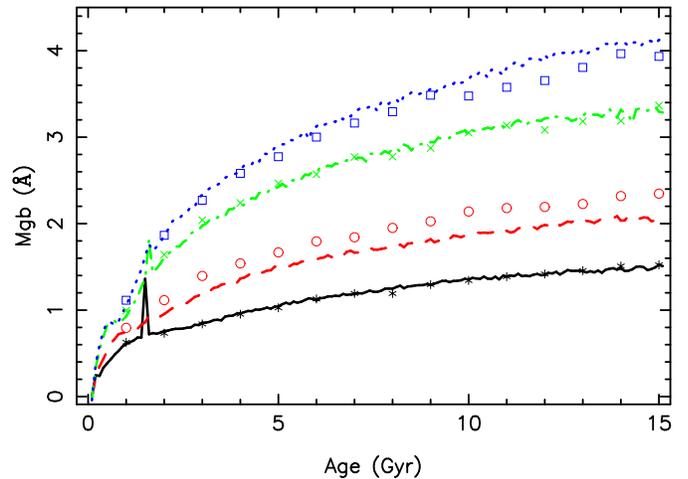}
  \caption{Similar to Fig. 7, but for Mgb.}
\end{figure}

\begin{figure}
  \includegraphics[angle=-90,width=88mm]{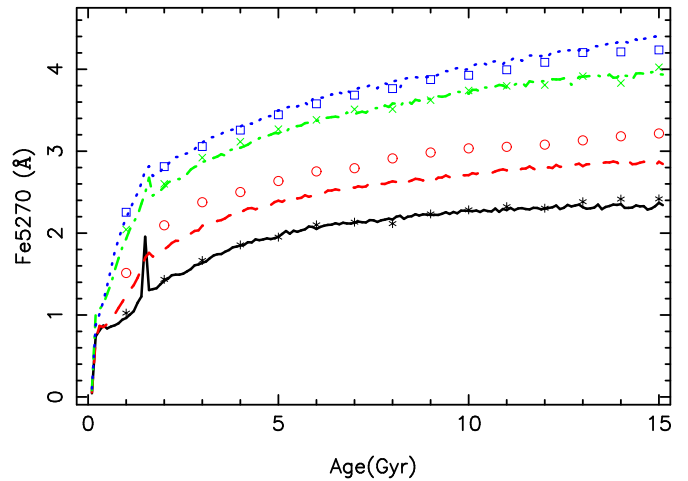}
  \caption{Similar to Fig. 7, but for Fe5270.}
\end{figure}

\begin{figure}
  \includegraphics[angle=-90,width=88mm]{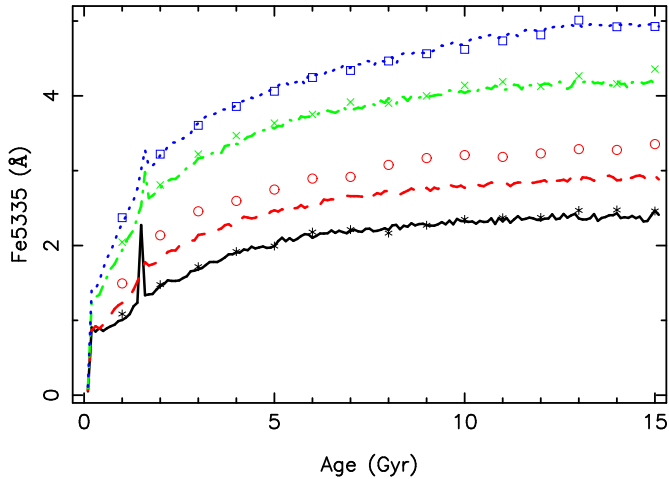}
  \caption{Similar to Fig. 7, but for Fe5335.}
\end{figure}

\subsection{Colour indices}
A few works have tried to model the populations of galaxies and star
clusters via bs-SSPs (e.g., \citealt{Zhang:2004, Zhang:2005}), but
there is no near-infrared colour indices presented. In fact, such
colours are very important for disentangling the well-known
degeneracy \citep{Worthey:1994} and useful for exploring the stellar
populations of distant galaxies (see, e.g.,
\citealt{Li:2007potential, Li:2007effects}). In addition, the
colours obtained by previous works (e.g., \citealt{Zhang:2004}) do
not evolve smoothly. In this work, we calculated some colours that
are potentially useful for stellar population studies. The BaSeL 3.1
spectra library \citep{Westera:2002} was used for calculating the
colours of populations because the wavelength coverage of Martins
library is only from 3\,000 to 7\,000 $\rm \AA$. Note that we take a
different method to calculate the colours of populations compared to
the works of \citet{Zhang:2004, Zhang:2005}. The colours in our work
are calculated by integrating SEDs, rather than by interpolating a
photometry library (see \citealt{Zhang:2004, Zhang:2005}).

\subsubsection{$UBVRIJHKLM$ colours}
The Johnson $UBVRIJHKLM$ colours are usually used for stellar
population studies. We calculated them for both bs-SSPs and ss-SSPs
with Salpeter and Chabrier IMFs. As an example, we plot the
evolution of four colours of bs-SSPs with Salpeter IMF in Fig. 11.
As we see, colours increase quickly with stellar age when the age is
less than about 2 Gyr and then they evolve slowly. The reason is
that the colours of young populations are dominated by massive
stars, which evolve very quickly before about 2 Gyr. On the other
hand, the colours of old ($>$ 2 Gyr) populations are mainly
dominated by less massive stars, which evolve more slowly than
massive stars. The figure shows that at fixed metallicity, colours
become redder with increasing age. At fixed stellar age, colours
become redder with increasing stellar metallicity. Therefore, the
stellar age and metallicity have similar effects on colours of
stellar populations. This is very the well-known age-metallicity
degeneracy. Therefore, it is impossible to disentangle the effects
of stellar age and metallicity completely. We can not give reliable
estimate for stellar age or metallicity using only one colour.
However, using a pair of colours, some constraints on the two
stellar-population parameters can be obtained, because the age and
metallicity sensitivities of each colour are usually different
\citep{Li:2007colourpairs}. Compared to the results of
\citet{Zhang:2004}, the colours calculated by this work evolve more
smoothly. We do not compare them here, as this can be understood
from the comparison of Lick indices. This is actually the first work
to calculate colours relating to near-infrared bands (e.g., $V-K$
and $I-K$) for bs-SSPs. Note that $(B-V)$ is sensitive to stellar
age while $(V-K)$ and $(I-K)$ to metallicity (see
\citealt{Li:2007colourpairs}).

\begin{figure}
  \includegraphics[angle=-90,width=88mm]{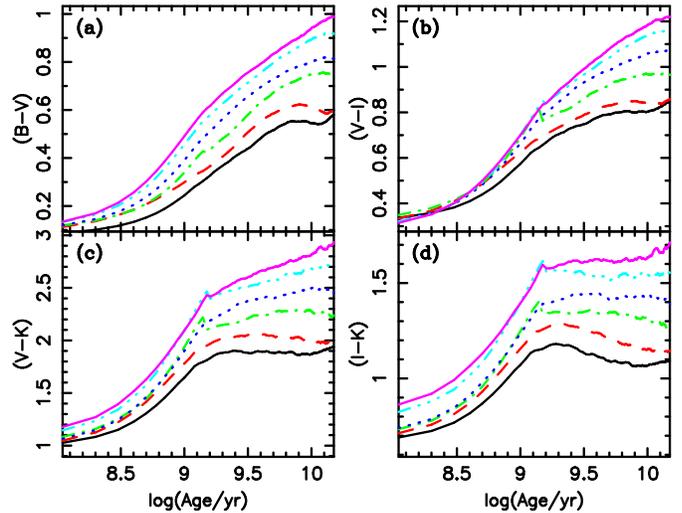}
  \caption{Evolution of four $UBVRIJHK$ colours of bs-SSPs with the Salpteter IMF.
 The bottom solid, dashed, dash-dotted, dotted, dash-dot-dot-dot and top solid lines
 are for metallicities of 0.0003, 0.001, 0.004, 0.01 and 0.02 and 0.03, respectively.}
\end{figure}

\subsubsection{$ugriz$ colours}
The Sloan Digital Sky Survey (SDSS) supplies a good deal of
observational data for astronomy studies and it takes an $ugriz$
system (hereafter SDSS-$ugriz$ system). In order to use its
photometry data for stellar population studies more conveniently, we
calculated some colours on the SDSS-$ugriz$ system. In Fig. 12, we
show the evolution of four colours. We see that SDSS-$ugriz$ colours
have similar peculiarities as $UBVRIJHKLM$ colours. Note that colour
$(u-r)$ is an age-sensitive colour and can be used to give
constraints on stellar-population parameters together with colours
such as $(r-K)$ \citep{Li:2007colourpairs}. This is the first work
to present the SDSS-$ugriz$ colours for bs-SSPs.

\begin{figure}
  \includegraphics[angle=-90,width=88mm]{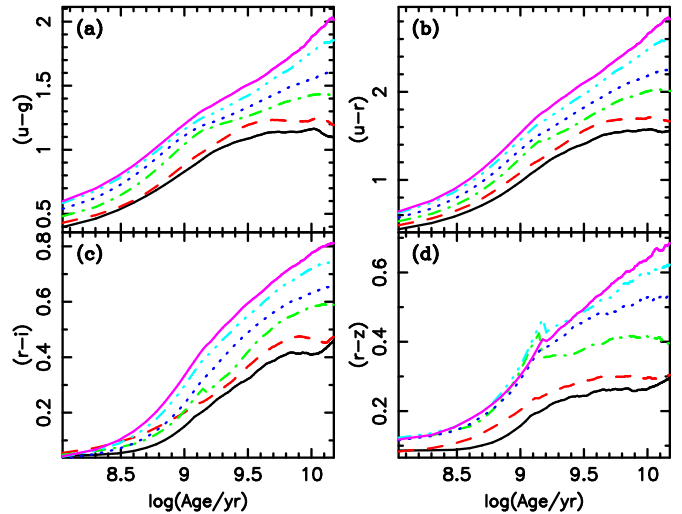}
  \caption{Similar to Fig. 11, but for four SDSS-$ugriz$ colours.}
\end{figure}

\subsubsection{Composite colours}
Because it is shown that some colours relating to both $UBVRIJHK$ and $ugriz$
magnitudes are more powerful than either $UBVRIJHK$ or $ugriz$ colours for
disentangling the stellar age--metallicity degeneracy \citep{Li:2007colourpairs},
we computed some such colours and call them \emph{composite colours}. In detail,
we calculated some colours consist of Johnson $UBVRIJHK$ magnitudes and SDSS-$ugriz$ magnitudes.
The evolution of four composite colours is shown in Fig. 13.
According to the work of \citet{Li:2007colourpairs}, $(u-R)$ and $(g-J)$
are sensitive to stellar age while $(r-K)$ and $(z-K)$ to stellar metallicity.

\begin{figure}
  \includegraphics[angle=-90,width=88mm]{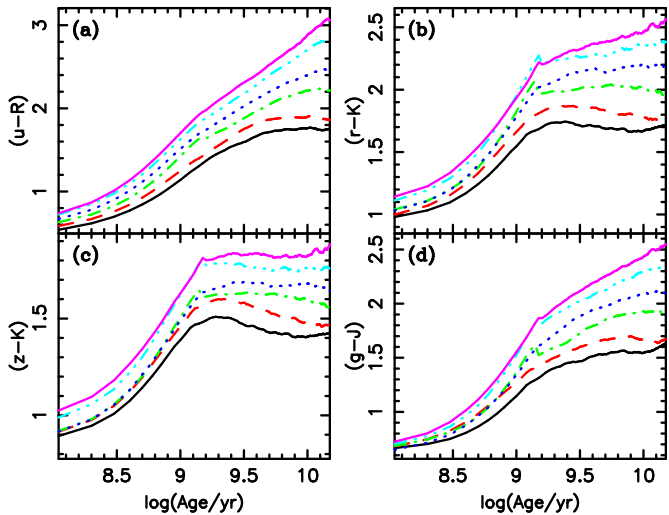}
  \caption{Similar to Fig. 11, but for four composite colours.}
\end{figure}

\subsubsection{Colour pairs for disentangling stellar age--metallicity degeneracy}
In this section, we show the color-color grids for two pairs of
colours (hereafter colour pairs) of bs-SSPs of our standard models.
The two pairs were shown to have the potential for disentangling the
age--metallicity degeneracy \citep{Li:2007colourpairs} via ss-SSPs
of \citet{Bruzual:2003}. They can possibly be used to separate the
stellar metallicity and age when bs-SSPs are used instead of
ss-SSPs. The two pairs are [$(u-R), (R-K)$], [$(u-R), (I-K)$]. The
detailed colour-colour grids of them can be seen in Figs. 14 and 15.
We find that the two colour-colour grids can disentangle the
age--metallicity degeneracy via bs-SSPs. Therefore, the two colour
pairs can be used for stellar population studies. However, the
uncertainties in stellar age and metallicity of metal-poor ($Z$ $<$
0.001) and old (Age $>$ 12 Gyr) populations seem larger than others.
Note that this two pairs are not necessarily the best pairs for
stellar population study. Some other potential pairs can be found in
our previous paper \citep{Li:2007colourpairs} and it is better to
choose colours according to the stellar-population peculiarities of
galaxies.

\begin{figure}
  \includegraphics[angle=-90,width=88mm]{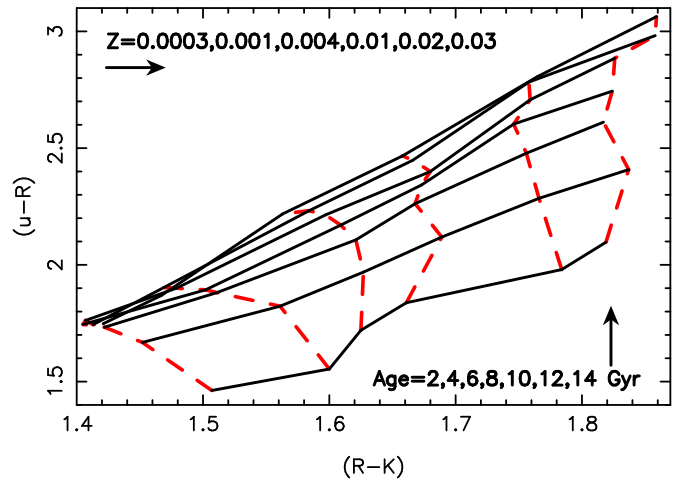}
  \caption{bs-SSP grids of $(u-R)$ versus $(R-K)$. The $u$ magnitude is on the SDSS-$ugriz$ system and $RK$ magnitudes are on the Johnson system.
Solid and dashed lines are for constant age and metallicity, respectively. }
\end{figure}

\begin{figure}
  \includegraphics[angle=-90,width=88mm]{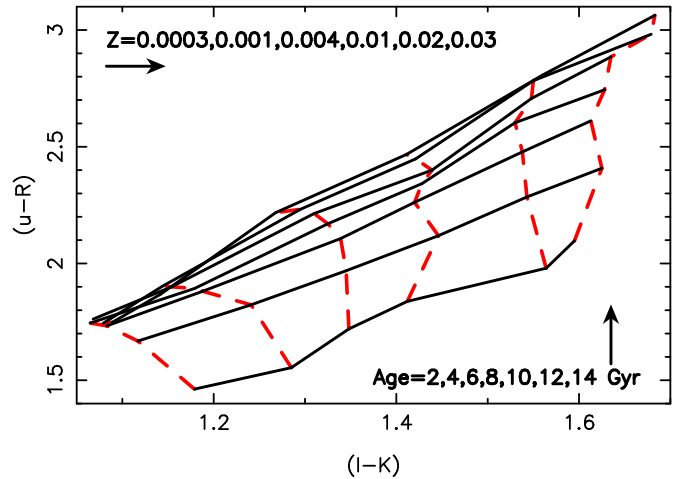}
  \caption{Similar to Fig. 14, but for $(u-R)$ versus $(I-K)$. The $I$ magnitude is on the Johnson system.}
\end{figure}

\subsection{Comparison with bs-SSP model and ss-SSP model}
Because most stellar population synthesis models are ss-SSP models
and they have been widely used for previous works, it is necessary
to compare the results (i.e., Lick indices and colours) of bs-SSP
model with those of ss-SSPs. We have a try in this work. We only
compare our results with those of the ss-SSPs of the model of
\citet{Bruzual:2003} (BC03 model) as most ss-SSP models gave similar
Lick indices and colours for the same stellar population (see, e.g.,
\citealt{Bruzual:2003}). The results can help us to understand how
the difference between the results obtained via ss-SSP models and
those obtained via bs-SSP models is. In Figs. 16 and 17, we compare
four widely used Lick indices and four colours of two kinds of SSPs.
Both the bs-SSP model and ss-SSP model take the Salpeter IMF and
calculate their colours via BaSeL 3.1 library, but the spectra
libraries used to calculate the Lick indices by two models are
various. Our bs-SSP model takes the Martins \citep{Martins:2005}
library and BC03 ss-SSP model mainly takes STELIB
\citep{LeBorgne:2003} library. Note that the Lick indices and
colours of BC03 ss-SSPs with metallicities of 0.01 and 0.03 were
interpolated because they were not given directly by the BC03 model.
Thus the values of the colours and Lick indices of BC03 ss-SSPs with
metallicities of 0.01 and 0.03 are possibly different from those
calculated by SEDs. From Fig. 16 we can find that a bs-SSP usually
has larger H$\beta$ index and less metal indices than a BC03 ss-SSP
when the two kinds of SSPs have the same stellar age and
metallicity. We are also shown that four colours of a bs-SSP are
bluer than those of an ss-SSP with the same age and metallicity.
Therefore, ss-SSP models will usually measure different values for
the stellar ages and metallicities of galaxies compared to bs-SSP
models. However, it seems that ss-SSP models and bs-SSP models can
give similar results for relative studies, because stellar age and
metallicity affect Lick indices and colours of the two kinds of
populations similarly.

\begin{figure*}
  \includegraphics[angle=-90,width=150mm]{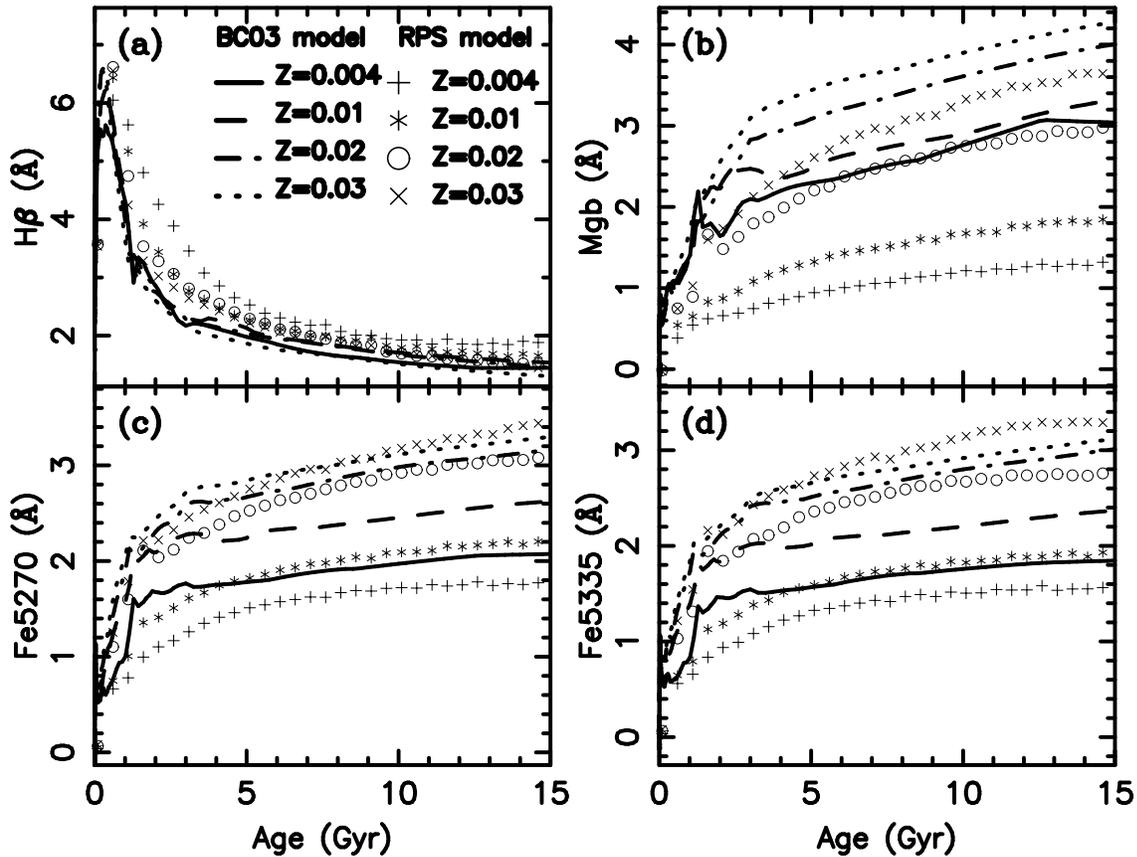}
  \caption{Comparison of four widely used Lick indices of bs-SSPs with those of BC03 ss-SSPs.
    Panels a), b), c) and d) are for H$\beta$, Mgb, Fe5270, and Fe5335, respectively.
    Lines and points in all panels have the same meanings as in Panel a).}
\end{figure*}

\begin{figure*}
  \includegraphics[angle=-90,width=150mm]{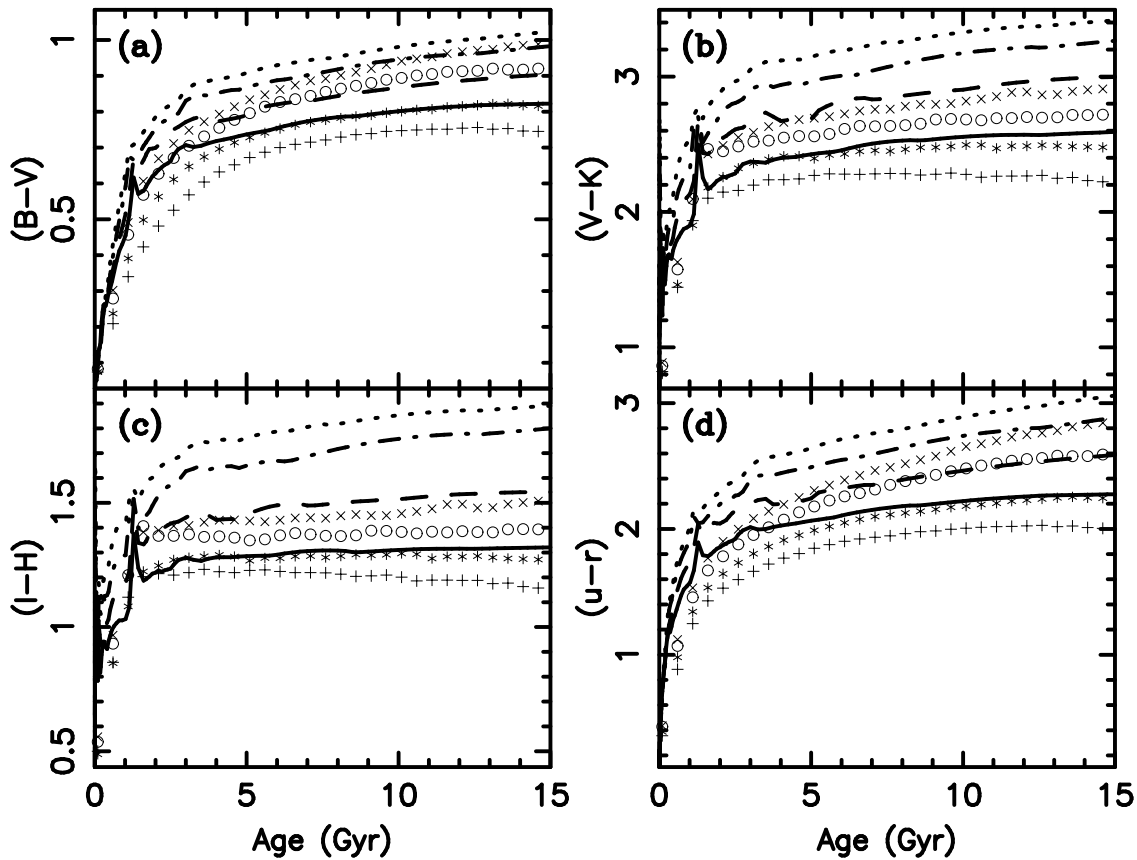}
  \caption{Comparison of four colours of bs-SSPs with those of BC03 ss-SSPs.
    Panels a), b), c) and d) are for $(B-V)$, $(V-K)$, $(I-H)$ and $(u-r)$, respectively.
    The $(u-r)$ colour is on the SDSS-$ugriz$ system, and the others are on the Johnson system.
    Lines and points have the same meanings as in Fig. 16.}
\end{figure*}

\section{CMD fitting for two star clusters}
Besides modeling stellar populations, the isochrone database can be
used for CMD fitting. Here we fitted the CMDs of star clusters M67
and NGC1868, via both bs-SSPs and ss-SSPs. The colours and
magnitudes were taken from the stellar population challenge of IAU
Symposium 241 (IAUS 241)
(http://www.astro.rug.nl/$^{\sim}$sctrager/challenge/) and the BaSeL
2.2 \citep{Lejeune:1997} spectra library was used to translate the
data of our isochrone database to CMDs of stellar populations. We
fitted the CMDs of the two above star clusters via stellar
populations with two IMFs. First, we fitted the CMDs via populations
with Salpeter IMF. As a result, via both bs-SSPs and ss-SSPs, star
cluster M67 was shown to have the solar metallicity and a 4.6 Gyr
age, with a distance modulus of 9.8 and a colour excess E(B-V) of
0.09. The stellar metallicity and age of NGC1868 are found to be
0.004 and 1.4 Gyr, respectively, with 18.8 for its distance modulus
and 0.04 for colour excess, E(V-I). Note that we only took stellar
populations with seven metallicities and 151 ages for our fittings,
according to the data of our isochrone database. In the work, CMDs
were fitted by comparing the theoretical and observational
percentage of stars in each sub-grid of the $gT$-grid compared to
the total number of stars in the observed colour and magnitude
ranges. Because stars after the turn-off point are very important
for determining stellar metallicity and the data of luminous stars
are more reliable than those of less luminous stars, the percentage
of stars in each sub-grid of $gT$-grid are weighted using their
luminosities. For convenience, we using 10$^{(V/-2.5)}$ instead of
the real luminosities of stars. Note that $gT$-grids of populations
were expressed by CMDs in the fitting. Here all stars are assumed to
be distinguishable, as the limitation of the database. However, it
is actually different from the true case, especially for the distant
star clusters. The fittings of the CMDs of M67 and NGC1868 are shown
in Fig. 18. Some sub-grids were not plotted because each of them
contains less than 0.5 luminosity-weighted stars and are not
important for fitting the CMDs when taking a luminosity-weighted fit
method. As we see, compared to ss-SSPs, bs-SSPs can fit some special
objects such as blue stragglers of M67 and as a whole, bs-SSPs
fitted the shapes of the CMDs of two star clusters better. This
suggests that the isochrones of bs-SSPs are closer to those of real
star clusters or galaxies. It seems that SSPs can fit the CMD of M67
well, but they fit the CMD of NGC1878 not so well. The best-fit CMD
of NGC1868 shows obviously different for the part with V magnitude
more luminous than about 22.5 mag. This possibly results from that
we assume each star of the star cluster can be distinguished but
actually many stars of NGC1868 were not distinguished in
observation. In fact, it is impossible to distinguish each star of
the star cluster, because the star cluster is too distant (with
distance modulus larger than 18). If some binary stars in the best
bs-SSP-fit population of NGC1868 can not be distinguished, the CMD
will be different from the one shown in Fig. 18, and it will be
possibly more similar to the observational CMD. When we fitted the
CMDs of the two star clusters via stellar populations with Chabrier
IMF, we got similar results. It implies that even if populations
with different assumptions (IMF, bs-SSP or ss-SSP) are used, we can
get reliable estimates for the stellar ages and metallicities of
star clusters via fitting CMDs. Note that the results of CMD fitting
is usually affected by the fitting method, thus the above results
are not always the best-fit results. In addition, because we only
use populations with seven metallicities to fit the populations of
the two star clusters, perhaps the results will change if we take
populations with more metallicities for fitting.

\begin{figure*}
  \includegraphics[angle=-90,width=176mm]{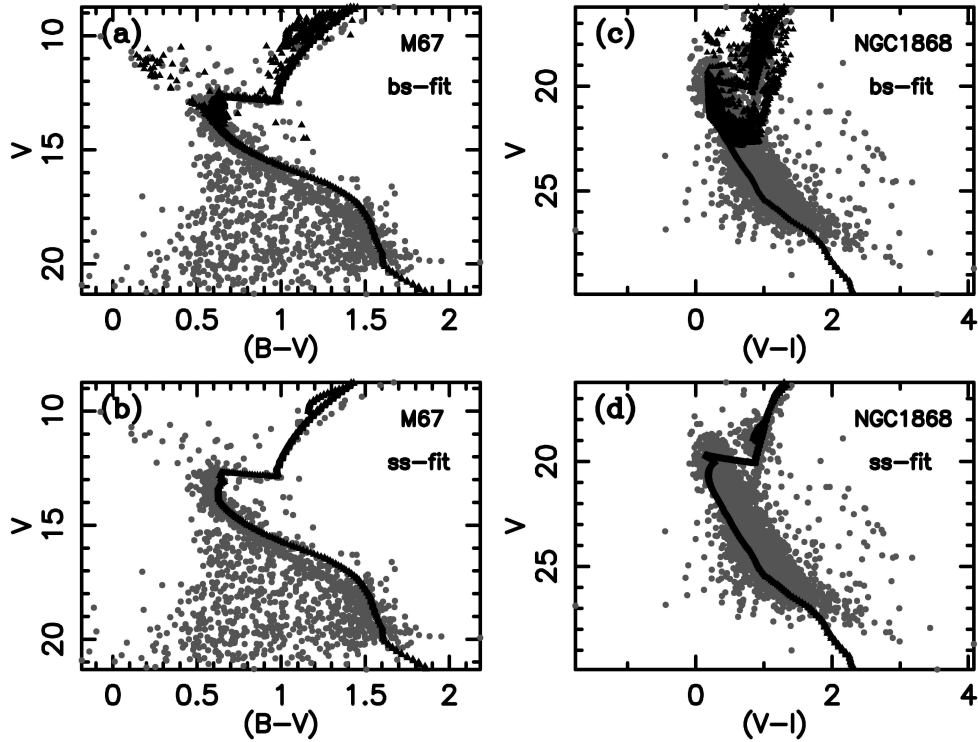}
  \caption{CMD fittings for star clusters M67 and NGC1868.
    Gray circles and dark triangles show the distributions of observational and theoretical stars, respectively.
    The upper and lower panels show the fittings via bs-SSPs and ss-SSPs, respectively. ``bs-fit'' and ``ss-fit'' mean
    the bs-SSP fitting and ss-SSP fitting, respectively.}
\end{figure*}

\section{discussions and conclusions}
We first presented an isochrone database for quickly modeling both
single star and binary star simple stellar populations (ss-SSPs and
bs-SSPs). The isochrone database only causes 0.81\% uncertainty in
stellar population synthesis results on average. Thus it can be
widely used for stellar population studies of galaxies and star
clusters. However, there are some points should be noted when
applying the database. First, because the star sample of populations
was generated via Monte Carlo technique, the stellar mass
distribution of a population is not continuous. As a result, some
line indices and colours possibly evolve with stellar age less
smoothly compared to those calculated via isochrone synthesis method
(see \citealt{Bruzual:2003}). This is more obvious for some
near-infrared colours. Second, the isochrone database can only model
stellar populations with metallicities from 0.0001 to 0.03, because
the Hurley code can not evolve stars with other metallicities. In
the work, we only calculated the isochrones of populations with
seven metallicities (0.0001, 0.0003, 0.001, 0.004, 0.01, 0.02 and
0.03), because stars with other metallicities can not be evolved as
well as stars with the above seven metallicities via Hurley code .
Furthermore, the distributions of the input parameters of binaries
can affect our results. In the work, we took a uniform distribution
for the mass ratio of a binary, $q$, as it is the widely used
distribution. However, the masses of the two components of a binary
perhaps are not correlated or the distribution of $q$ is not a
uniform distribution. If it is that case, we will possibly get some
various results. In addition, we only present the isochrone database
for two widely used IMFs, as the limitation of our computing
ability. There are actually some other types for the IMF of stellar
populations. We intend to study further in the future. Although we
mainly aim to model simple stellar populations (bs-SSPs and
ss-SSPs), the isochrone database can be used to model composite
stellar populations (CSPs). The database makes it easier to study
the spectra and photometry evolution of galaxies via bs-SSPs. This
is useful for future studies on the formation and evolution of
galaxies. It can also be used to study the effects of binary
interactions on stellar population synthesis studies. In fact, to
investigate how binary interactions affect the results of stellar
population studies is very the topic of another work.

Then we introduced a rapid stellar population synthesis ($RPS$)
model based on our isochrone database. The $RPS$ model calculated
high-resolution (0.3 $\rm \AA$) spectral energy distributions
(SEDs), Lick indices, and colour indices of bs-SSPs and ss-SSPs, for
two widely used IMFs. However, the SEDs and Lick indices are
available only for populations with metallicities of 0.004, 0.01,
0.02 and 0.03. The reason is that the metallicity coverage of the
spectra library used for calculating SEDs of stellar populations is
0.002--0.04. Therefore, the metallicity range of theoretical
populations is possibly not wide enough for some studies. However,
$UBVRIJHK$ colours, SDSS-$ugriz$ colours and composite colours are
available for populations with metallicities from 0.0003 to 0.03,
which can be used to study the stellar-population parameters (age
and metallicity) of most galaxies. When we compared four widely used
Lick indices and four colours of bs-SSPs of RPS model to those of
ss-SSPs of the BC03 model (\citealt{Bruzual:2003}, BC03), we found
that some various results will be shown if we take bs-SSPs instead
of BC03 ss-SSPs for stellar population studies. However, it seems
that even if bs-SSPs are used instead of ss-SSPs for studies, the
relative values of stellar populations will not change obviously, as
stellar age and metallicity affect the Lick indices and colours of
bs-SSPs and ss-SSPs similarly.

Because the isochrone database can also be used to fit the
colour-magnitude diagrams (CMDs) of star clusters, we fitted the
CMDs of two star clusters (M67 and NGC1868) and gave their stellar
metallicities, ages, distances and colour excesses, as examples. Our
results showed that bs-SSPs can fit some special stars such as blue
stragglers of star clusters, which can not be fitted by ss-SSPs.
This suggests that one should consider binary interactions in
stellar population synthesis studies. However, because the database
assumes that all stars in a stellar population are distinguishable
and some stars of distant star clusters can usually not be
distinguished, the database is more suitable to fit the CMDs of
nearby star clusters. When the database is used to fit the CMDs of
very distant galaxies, one should take the above point into account.

\section*{Acknowledgments}

We thank a reviewer of MNRAS, Profs. Tinggui Wan, Hongyan Zhou,
Fenghui Zhang and Drs. Guoliang L\"{u}, Xiangcun Meng for useful
discussions and Prof. Scott Trager for the data of two star
clusters. This work is supported by the Chinese National Science
Foundation (Grant Nos 10433030, 10521001 and 2007CB815406), and the
Chinese Academy of Science (No. KJX2-SW-T06).

\bsp

\label{lastpage}

\end{document}